# RCSB PDB AI Help Desk: retrieval-augmented generation for protein structure deposition support

Vivek Reddy Chithari[1], Jasmine Y. Young[1], Irina Persikova[1], Yuhe Liang[1], Gregg V. Crichlow[1], Justin W. Flatt[1], Sutapa Ghosh[1], Brian P. Hudson[1], Ezra Peisach[1], Monica Sekharan[1], Chenghua Shao[1], and Stephen K. Burley[1,2]

[1] *RCSB Protein Data Bank, Rutgers, The State University of New Jersey, Piscataway, NJ, USA*

[2] *RCSB Protein Data Bank, San Diego Supercomputer Center, University of California San Diego, CA, USA*

## Abstract

**Motivation:** Structural Biologists have contributed more than 245,000 experimentally determined three-dimensional structures of biological macromolecules to the Protein Data Bank (PDB). Incoming data are validated and biocurated by ~20 expert biocurators across the wwPDB. RCSB PDB biocurators who process more than 40% of global depositions face increasing challenges in maintaining efficient Help Desk operations, with approximately 19,000 messages in approximately 8,000 entries received from depositors in 2025.

**Results:** We developed an AI-powered Help Desk using Retrieval-Augmented Generation (RAG) built on LangChain with a pgvector store (PostgreSQL) and GPT-4.1-mini. The system employs pymupdf4llm for Markdown-preserving PDF extraction, two-stage document chunking, Maximal Marginal Relevance retrieval, a topical guardrail that filters off-topic queries, and a specialized system prompt that prevents exposure of internal terminology. A dual-LLM architecture uses separate model configurations for question condensing and response generation. Deployed in production on Kubernetes with PostgreSQL (pgvector), it provides around-the-clock depositor assistance with citation-backed, streaming responses.

**Availability and implementation:** Freely available at https://rcsb-deposit-help.rcsb.org.

## 1 Introduction

The Protein Data Bank (PDB) (Protein Data Bank, 1971; Berman et al., 2000; wwPDB Consortium, 2019) was established in 1971 as the first open-access digital data resource in biology, initially comprising just seven X-ray crystal structures of proteins. Today, the PDB archive houses more than 245,000 experimentally determined, three-dimensional atomic-level structures of biological macromolecules that are freely accessed and used by millions of researchers, educators, and students worldwide. This wealth of 3D biostructure information serves as a cornerstone for basic and applied research across fundamental biology, biomedicine, biotechnology, bioengineering, and energy sciences.

The archive continues to grow, with depositions increasing in number, size, and complexity as new experimental methods emerge and existing techniques are refined.

The RCSB Protein Data Bank (RCSB PDB) (Burley et al., 2019) serves as one of the founding members of the Worldwide Protein Data Bank (wwPDB) (wwPDB Consortium, 2019), which collectively manages the PDB archive. The RCSB PDB, which serves as the PDB Core Archive Keeper designated by the wwPDB, is responsible for packaging and disseminating PDB data to the public, ensuring that structural data are freely and publicly available to the global research community. RCSB PDB provides essential services including data deposition, validation, and biocuration of data coming from the Americas, along with dissemination of global macromolecular structure data to the PDB Core Archive. Central to these operations is the biocuration process (Young et al., 2018), wherein expert human biocurators review deposited structures to ensure data quality, consistency, and adherence to community standards before public release.

wwPDB surpassed the milestone of 20,000 new structure depositions in 2025 alone. This achievement reflects the impact of rapid advances in cryo-electron microscopy, high-throughput crystallography, enhanced computational and AI-driven modeling tools, and a long-standing commitment to open data sharing. With the sustained growth in PDB depositions and federal funding that does not keep pace with inflation, RCSB PDB biocurators face mounting operational challenges. Depositors, *i.e.*, structural biology researchers submitting their data to the archive, frequently require assistance (RCSB PDB received ~19,000 messages in ~8000 entries from depositors in 2025) navigating deposition procedures, understanding validation reports, and resolving technical issues. Historically, this support has been provided through a human-staffed Help Desk by the wwPDB biocurators, via OneDep User Interface (Young et al., 2017) or emails. This represents one of the most time-consuming daily tasks carried out by RCSB PDB biocurators. Depositor queries span a wide range of topics from basic procedural questions to complex technical issues requiring detailed explanations drawn from extensive wwPDB documentation.

The emergence of large language models (LLMs) and Retrieval-Augmented Generation (RAG) (Lewis et al., 2020) architectures presents an opportunity to enhance Help Desk operations through intelligent automation. RAG systems combine the generative capabilities of LLMs with targeted information retrieval from domain-specific knowledge bases, enabling contextually accurate responses grounded in authoritative source material. Unlike purely generative approaches that may produce plausible but inaccurate information, a phenomenon known as "hallucination", RAG architectures retrieve relevant passages from curated documents before generating responses, substantially improving factual accuracy and enabling transparent citation of sources.

Herein, we present an AI-powered Help Desk (https://rcsb-deposit-help.rcsb.org) designed to assist PDB depositors with queries pertaining to 3D biostructure data deposition, validation, and biocuration.

The system leverages a LangChain-based (Chase, 2026) RAG architecture with a pgvector (pgvector, 2026) vector store and OpenAI language models to provide contextual, actionable responses drawn from authoritative wwPDB documentation. We describe the technical implementation, document-processing pipeline, and integration strategies that enable effective deployment as a public-facing help resource. This work demonstrates how AI applications can be thoughtfully applied to improve biocuration efficiency while maintaining the high standards of accuracy and transparency expected in scientific data management.

## 2 System Architecture and Implementation

### 2.1 Overall Architecture

The RCSB PDB AI Help Desk employs a modular four-layer architecture comprising: (i) a presentation layer for user interaction, (ii) an application layer for request handling and session management, (iii) an inference layer implementing the RAG pipeline, and (iv) a knowledge base layer for document storage and semantic retrieval. This separation of concerns enables independent scaling and maintenance of each component while facilitating integration with existing RCSB PDB infrastructure.

The presentation layer delivers a web-based chat interface built with plain HTML, CSS, and JavaScript, using marked.js for Markdown rendering and DOMPurify for output sanitization. Users interact with the system through a conversational paradigm, submitting natural language queries and receiving formatted responses with embedded source citations. The interface is designed for anonymous access without authentication requirements, functioning as a public helpdesk resource where depositors can seek immediate assistance without creating accounts or maintaining persistent sessions. This design philosophy prioritizes accessibility and ease of use, lowering barriers for depositors who may have urgent questions during the deposition process.

The application layer is built on FastAPI with uvicorn, providing REST endpoints for session management and Server-Sent Events (SSE) for streaming responses. Session persistence uses PostgreSQL with a dual-pool architecture: a psycopg connection pool for session operations and a separate SQLAlchemy-based pool for vector store queries, providing for concurrent access safety. The layer incorporates rate limiting via SlowAPI (10 requests per minute for chat and session endpoints, 20 requests per minute for feedback) and security middleware enforcing Content Security Policy (CSP), HTTP Strict Transport Security (HSTS), and request size limits (64 KB body limit, 10,000-character message limit). A per-session cap of 50 chat threads prevents resource exhaustion.

### 2.2 Knowledge Base Construction

The foundation of any RAG system (Lewis et al., 2020) is its knowledge base, which must accurately represent the source documentation while enabling efficient semantic retrieval. For the RCSB PDB AI Help Desk, we constructed a specialized knowledge base from authoritative wwPDB documentation, including deposition policies, biocuration procedures, validation guidelines, and frequently asked questions. The knowledge base encompasses documentation pertaining to the full lifecycle of structure data, from initial submission through validation, annotation, and public release.

Document processing employs pymupdf4llm, a Markdown-aware PDF extraction library that converts PDF content into structured Markdown text. Scientific documentation frequently contains structured elements such as tables, multi-column layouts, headers, bold text, lists, and embedded figures that challenge conventional text extraction approaches. Unlike plain-text extractors, pymupdf4llm preserves document formatting as Markdown syntax, maintaining headers, bold emphasis, list structures, and table layouts. This capability is particularly important for wwPDB documentation, which often presents procedural information in tabulated formats or uses visual organization to convey relationships between concepts.

Following document parsing, content is segmented into retrievable chunks through a two-stage splitting process optimized for the RAG retrieval pipeline. The first stage applies LangChain's MarkdownTextSplitter, which respects Markdown structural boundaries such as headers and sections, ensuring that chunks align with the logical organization of the source documents. The second stage applies RecursiveCharacterTextSplitter with a chunk size of 2,000 characters and 400-character overlap (20%), using a hierarchy of separators (paragraph breaks "\n\n", line breaks "\n", sentence boundaries ". ", and word boundaries " ") to preserve semantic coherence within chunks. This two-stage approach balances the competing requirements of maintaining sufficient context within each chunk while enabling precise retrieval of relevant passages.

Document chunks are embedded using OpenAI's text-embedding-3-small model (OpenAI, 2024) and stored in a pgvector-enabled PostgreSQL database (pgvector, 2026), queried using cosine distance for similarity matching. The ingestion pipeline implements a shadow table swap strategy for zero-downtime updates: new embeddings are written to a staging table, then atomically swapped with the production table, ensuring uninterrupted service during knowledge base rebuilds. The ingestion pipeline implements change detection through a SHA-256 manifest system: file hashes are computed and compared against a stored manifest to detect additions, modifications, and deletions, triggering a full rebuild only when content changes are detected. An optional Google Drive integration enables automated synchronization of documentation updates from shared organizational folders.

## 2.3 Language Model Integration

The system employs a dual-LLM architecture with two distinct GPT-4.1-mini (OpenAI, 2025) instances, each configured each serving specific roles in the system. Document chunks are encoded using OpenAI's text-embedding-3-small (OpenAI, 2024), an embedding model optimized for semantic similarity matching. These embeddings enable dense retrieval, wherein user queries are similarly encoded and matched against the document corpus using cosine similarity metrics.

The primary QA LLM handles response generation and is configured with temperature 0.0 (fully deterministic) and a maximum token limit of 8,192, with streaming enabled for progressive response delivery. The conservative temperature setting minimizes stochastic variation in outputs, prioritizing factual accuracy over creative generation. A separate Condense LLM handles question condensing for multi-turn conversations, configured with temperature 0 (fully deterministic) and a maximum token limit of 512, operating in non-streaming mode for efficiency. This separation ensures that follow-up questions are resolved into standalone queries with maximum consistency before retrieval.

Retrieved context is injected into the system prompt rather than the human message turn, following a context-in-system pattern that keeps the user's question cleanly separated from the reference material. These conservative generation parameters, combined with the grounding provided by retrieved context, substantially reduce the risk of hallucinated content. The model is prompted to synthesize information from retrieved passages while maintaining fidelity to source material and explicitly citing the documents from which information is drawn.

## 2.4 System Prompt Engineering

A critical aspect of deploying the AI Help Desk for public-facing use involves careful engineering of the system prompt governing model behavior. The system prompt establishes the assistant's persona, defines response formatting conventions, and specifies constraints on content generation. For the RCSB PDB context, the system prompt was iteratively refined to eliminate internal biocuration terminology that would be inappropriate or confusing for external depositors.

The knowledge base includes internal documentation that references biocurator workflows, internal systems, and staff-specific procedures. While this content provides valuable context for understanding deposition processes, direct exposure of such terminology could confuse depositors or inappropriately reveal internal operational details. The system prompt explicitly instructs the model to reformulate responses in depositor-appropriate language, translating internal references into externally meaningful guidance. For example, references to internal annotation systems are reformulated as guidance about what depositors should expect during the biocuration process without exposing system-specific terminology.

The system prompt is structured around seven numbered behavioral guidelines that govern response generation. A FORBIDDEN CONTENT block enumerates eleven categories of prohibited content, including internal staff names, biocurator-specific jargon, triage terminology, and internal system identifiers, including email addresses in responses (directing users to the feedback section instead) and responding to off-topic messages as if they were legitimate inquiries, ensuring that none of these items appear in depositor-facing responses. A complementary REQUIRED APPROACH block specifies seven directives covering depositor-appropriate language reformulation, citation formatting, knowledge base boundary enforcement, and solution prioritization rules that lead with the most actionable guidance. When a query falls outside the scope of the knowledge base, the system prompt directs the model to acknowledge the limitation and suggest that users provide feedback via the rating and comments section below the response to reach the support team.

Response formatting conventions specified in the system prompt ensure consistent, readable outputs. The model is instructed to provide direct answers to queries, cite relevant documentation, and offer step-by-step procedures where applicable without excessive verbosity. Responses are structured to prioritize actionable guidance, enabling depositors to quickly understand what steps they should take to address their concerns. These constraints were iteratively refined based on biocurator feedback during internal testing.

# 3 Retrieval-Augmented Generation Pipeline

## 3.1 Query Processing and Retrieval

When a user submits a query, the system initiates a multi-stage retrieval and generation pipeline. For multi-turn conversations, the system first condenses follow-up questions into standalone queries using the dedicated zero-temperature Condense LLM that incorporates the previous conversation context. Conversation history is maintained in an in-memory Least Recently Used (LRU) cache supporting up to 3 turns per chat thread, with a capacity of 500 chat threads before least-recently-used eviction. This condensation step ensures that pronoun references and implicit context from prior exchanges are resolved before retrieval.

λ=0.7 (relevance-diversity trade-off parameter favoring relevance). Query embeddings are matched against the pgvector store using cosine distance indexing. The MMR strategy ensures that the language model receives diverse, complementary context rather than multiple near-duplicate passages that would waste the limited context window.

The condensed query is then encoded using the same text-embedding-3-small model applied to document chunks, producing a dense vector representation that captures semantic meaning. Document retrieval employs Maximal Marginal Relevance (MMR) search (Carbonell and Goldstein, 1998), which balances relevance to the query against diversity among retrieved passages to reduce redundancy. The retriever is configured with k=8 (number of documents returned), fetch_k=30 (candidate pool size), and λ=0.7 (relevance-diversity trade-off parameter favoring relevance). Query embeddings are matched against the pgvector store using cosine distance indexing. The MMR strategy ensures that the language model receives diverse, complementary context rather than multiple near-duplicate passages that would waste the limited context window.

## 3.2 Response Generation with Citations

Retrieved passages are injected into the system prompt along with the carefully engineered behavioral constraints, forming the complete context provided to GPT-4.1-mini (OpenAI, 2025) for response generation. The model is instructed to synthesize a response that directly addresses the user's question while grounding claims in the retrieved content. Importantly, the model is required to cite the specific documents from which information is drawn, enabling users to verify responses against authoritative sources.

Citations are rendered inline within responses, identifying the source document for each substantive claim. This transparency is essential for scientific applications, where users must be able to assess the provenance and authority of information. Users who require additional detail beyond the synthesized response can consult the cited documentation directly, supporting a seamless transition from AI-assisted guidance to in-depth review of source material.

Response generation employs streaming output via Server-Sent Events (SSE) using the sse-starlette library, displaying text progressively as it is generated rather than waiting for complete response synthesis. Each SSE chunk transmits the accumulated content generated so far rather than individual deltas, simplifying client-side rendering. A 5-minute timeout accommodates longer, multi-source responses. The SSE implementation includes HAProxy-resilient flush padding (three flush comments after done or error events) to ensure reliable delivery through the Kubernetes ingress layer. Streaming is particularly valuable for longer responses that synthesize information from multiple sources, providing immediate feedback to users and reducing perceived latency.

### 3.3 Handling Out-of-Scope Queries

Not all user queries will be addressable from the knowledge base. Queries about general scientific concepts, specific experimental techniques outside the deposition context, or topics entirely unrelated to PDB operations require appropriate handling to maintain user trust. The system is configured to acknowledge when a query falls outside the scope of available documentation, directing users to alternative resources or human assistance as appropriate. This honest acknowledgment of limitations is preferable to generating speculative responses that could mislead users.

Prior to retrieval, the system employs a topical guardrail that classifies incoming messages as on-topic or off-topic using a lightweight LLM call (reusing the Condense LLM instance with temperature 0). The guardrail permits queries related to PDB/EMDB structure deposition, validation reports, biocuration, wwPDB policies, data formats, structural biology methods, and general greetings, while declining sales pitches, vendor registrations, job applications, and unrelated spam. This pre-retrieval filter prevents the RAG pipeline from wasting computation on irrelevant queries and ensures appropriate responses for off-topic messages.

Out-of-scope query handling operates at two levels. The topical guardrail described above filters clearly off-topic messages before retrieval. Additionally, the system prompt instructs the language model to acknowledge when retrieved passages are insufficient to answer a query, directing users to the feedback section to reach the support team for queries requiring personalized assistance or information not covered by existing documentation.

## 4 Deployment and Integration

### 4.1 Containerized Deployment

The AI Help Desk is deployed using Docker (Merkel, 2014) containerization, enabling consistent operation across development, testing, and production environments. The container is built from a python:3.12-slim base image using a multi-stage build with the uv package manager for fast, reproducible dependency resolution. Containerization encapsulates all system dependencies, ensuring that the application behaves identically regardless of the underlying host infrastructure. This approach simplifies deployment and maintenance while enabling horizontal scaling through the replication of container instances to accommodate increased workload.

The deployment architecture packages the FastAPI application serving both REST and SSE endpoints into a single container, eliminating inter-service communication overhead. The system is deployed on Kubernetes via a Helm chart with HAProxy ingress, CephFS ReadWriteMany Persistent Volume Claims for shared storage, and ExternalSecrets for secure API key management. A CloudNative PostgreSQL (CNPG) cluster with pgvector extension provides both the vector store and session persistence, managed as a separate Kubernetes resource with primary and read-only replica instances. An admin dashboard runs as a separate pod for feedback analysis, isolated from RAG/LLM dependencies. A daily CronJob backs up the PostgreSQL session database to Google Drive with 30-day retention. Automated CI/CD through GitHub Actions handles build, push to the container registry, and rolling deployment. Health checks at the /api/health endpoint enable Kubernetes liveness and readiness probes.

The system implements manifest-based hot-reload for the vector store, enabling knowledge base updates without pod restarts. When source documents change, the SHA-256 manifest detects modifications and triggers a full knowledge base rebuild. The ingestion pipeline uses a shadow table swap strategy for zero-downtime updates, writing new embeddings to a staging table and atomically swapping it with the production table. Error recovery logic catches database connection errors, transparently reloading the vector store to maintain continuous availability. If the vector store is unavailable at startup, the application enters a degraded mode and automatically recovers when the pgvector table becomes available, checked via health probe retries within approximately 10 seconds. A two-transaction commit strategy ensures reliability during streaming: the user message is committed to the session database before streaming begins, and the assistant response is committed after completion, preventing data loss if streaming is interrupted.

### 4.2 Knowledge Base Maintenance

RCSB PDB AI Help Desk

Maintaining currency of the knowledge base is essential for ensuring that responses reflect the latest policies, procedures, and guidance. The system implements a change-detection mechanism based on SHA-256 file hashing that determines whether a full knowledge base rebuild is needed. When source documentation changes are detected, the entire vector store is rebuilt using the shadow table swap strategy for zero-downtime updates. When no changes are detected, the rebuild is skipped entirely, avoiding unnecessary reprocessing.

When any source document changes, the synchronization process detects the modification via SHA-256 hashing and triggers a full knowledge base rebuild. The shadow table swap strategy ensures this rebuild occurs with zero downtime, maintaining continuous service availability during reindexing.

An optional Google Drive integration enables automated synchronization of documentation updates from shared organizational folders, scheduled to run every fifteen minutes via a Kubernetes CronJob. A separate weekly CronJob exports accumulated user feedback data to Google Drive for analysis, enabling systematic review of query patterns and response quality. The feedback schema captures the question, a 200-character answer preview, full answer length, number of references cited, and a star rating, providing structured data for continuous improvement.

## 5 Results and Observations

### 5.1 Knowledge Base Statistics

The current knowledge base encompasses core wwPDB documentation including deposition policies, biocuration procedures, validation FAQs, and EMDB policies. Document processing with pymupdf4llm extraction and two-stage chunking (MarkdownTextSplitter followed by RecursiveCharacterTextSplitter) produces a corpus of 183 chunks totaling approximately 92,000 tokens. The chunk distribution reflects the relative volume and complexity of each documentation source, with procedural documents contributing the largest share of chunks due to their detailed step-by-step content.

All source documents have been successfully ingested and processed with a 100% success rate, based on completion of the ingestion pipeline without errors. Processing time for the complete knowledge base initialization is under two minutes, enabling rapid deployment and iteration during development. The SHA-256 change-detection mechanism avoids unnecessary full rebuilds when documentation has not changed, supporting efficient maintenance as documentation evolves.

### 5.2 Response Quality Assessment

Preliminary internal evaluation of the AI Help Desk demonstrates effective handling of common depositor queries. The system successfully addresses questions about deposition procedures, validation report interpretation, file format requirements, and policy clarifications. Responses consistently cite relevant source documents, enabling verification of provided information. The streaming response mechanism provides satisfactory user experience, with initial content appearing within seconds of query submission.

Internal testing identified several areas for refinement that were addressed before public release. Initial implementations occasionally exposed internal biocurator terminology in responses; system prompt refinements successfully eliminated such leakage. Reference formatting required optimization to ensure consistent, readable citation presentation. Documentation updates were required in some cases where source material contained outdated information (such as deprecated URLs), highlighting the importance of knowledge base maintenance procedures.

### 5.3 Coverage and Limitations

The AI Help Desk is designed to address queries that can be answered from existing wwPDB documentation. This scope encompasses the majority of procedural and policy questions that our depositors typically submit to the human Help Desk. However, queries requiring access to specific deposition records, personalized account assistance, or information not documented in the knowledge base remain outside the system's capabilities. For such queries, the system appropriately directs users to contact the human Help Desk.

The conversational interface supports multi-turn dialogue, enabling users to ask follow-up questions that refine or extend initial queries. Conversation context is maintained within a session, allowing the system to interpret subsequent questions in light of previous exchanges. This capability supports natural exploration of complex topics that may require multiple rounds of clarification.

## 6 Discussion

### 6.1 Implications for Depositors and Biocuration Operations

The AI Help Desk ([https://rcsb-deposit-help.rcsb.org](https://rcsb-deposit-help.rcsb.org)) represents a significant step toward applying artificial intelligence to enhance biocuration operations at RCSB PDB. By automating responses to routine documentation queries, the system has the potential to reduce the burden on human biocurators, freeing their time for tasks requiring expert judgment that cannot be relegated to automated systems. The persistence of source citations in AI-generated responses maintains the transparency and verifiability that are essential in scientific data management contexts.

It is important to emphasize that the AI Help Desk is designed to complement, not replace, human expertise. Complex queries requiring interpretation of specific experimental data, judgment calls about unusual deposition scenarios, or issues requiring access to internal systems will continue to require human intervention. The AI system serves as a first-line resource that can address straightforward queries immediately, escalating to human biocurators only when necessary. This tiered approach optimizes the allocation of expert human attention while ensuring that depositors receive timely assistance.

From the depositor perspective, the AI Help Desk offers immediate, around-the-clock access to guidance on deposition procedures, validation report interpretation, and policy clarifications — without the delays inherent in waiting for human responses. Depositors can resolve routine questions independently at any stage of the deposition workflow, reducing friction and accelerating the path from submission to public release. It also reduces the effort made by depositors on looking up various documentation to find relevant information. The citation-backed responses further empower depositors to explore relevant documentation on their own, fostering greater familiarity with wwPDB standards and potentially improving the quality of initial submissions over time.

**6.2 Technical Considerations**

The choice of RAG architecture over fine-tuned models offers several advantages for this application. RAG systems can be updated simply by modifying the knowledge base, without requiring expensive model retraining. This flexibility is crucial for a domain where policies and procedures evolve regularly. Additionally, the explicit retrieval step provides interpretability that pure generative models lack—users can see exactly which documents informed a response, supporting verification and trust. The MMR retrieval strategy further improves response quality by reducing redundancy among retrieved passages, ensuring that the language model receives diverse, complementary context.

The integration of pymupdf4llm for Markdown-preserving PDF extraction and two-stage chunking (MarkdownTextSplitter followed by RecursiveCharacterTextSplitter) reflects the specialized requirements of scientific documentation. Generic text extraction approaches would fail to preserve the structural information encoded in tables, lists, and formatted layouts that convey important meaning in procedural documents. The two-stage splitting strategy first respects Markdown structural boundaries and then enforces consistent chunk sizes, preserving both logical document organization and fine-grained retrievability.

The two-transaction commit strategy ensures data reliability during streaming: user messages are persisted before generation begins and assistant responses are committed after completion, preventing data loss from interrupted streams. Automatic error recovery—catching database connection errors and transparently reloading the vector store—provides continuous availability without manual intervention, an important consideration for a production system serving depositors across time zones.

**6.3 Future Directions**

Several directions for future development are envisioned. Expansion of the knowledge base to encompass additional documentation sources will broaden the range of queries the system can address. Integration with RCSB PDB web services could enable the system to provide dynamic information about specific entries or depositions, moving beyond static documentation to real-time data access. User feedback mechanisms will support continuous improvement, enabling identification of queries that are poorly handled and documentation gaps that should be addressed.

In the longer-term, AI assistance could extend to support PDB data users on the Help Desk and/or beyond the Help Desk to support biocurators directly in their annotation work. Intelligent systems could suggest annotations, flag potential issues, or retrieve relevant precedents from previously curated entries. Such applications would require careful design to ensure that AI assistance enhances rather than undermines the expert judgment that defines high-quality biocuration.

# 7 Conclusion

We have presented an AI-powered Help Desk prototype for RCSB PDB that leverages Retrieval-Augmented Generation to provide contextual, citation-backed responses to depositor queries. The system demonstrates how modern AI techniques can be applied to reduce operational burden while maintaining the transparency and accuracy essential for scientific data management. By grounding responses in authoritative documentation and providing explicit citations, the system supports rather than supplants human expertise.

The prototype addresses a genuine operational challenge—the time-intensive nature of Help Desk support—while respecting the constraints of a scientific context that demands accuracy and verifiability. The technical implementation demonstrates effective integration of Markdown-preserving document extraction, two-stage chunking, MMR retrieval, and large language model capabilities, a topical guardrail for off-topic filtering, and a shadow table swap strategy for zero-downtime knowledge base updates within a modular, maintainable four-layer architecture. This work establishes a foundation for broader application of AI techniques to support biocuration operations at RCSB PDB and potentially other scientific data resources facing similar challenges.

## Acknowledgements

We thank Rachel Green for providing a general help desk to PDB users and depositors, Aditya Pingale for the IT support on this work, and wwPDB biocurators for deposition support globally. AI-assisted tools (GitHub Copilot and ChatGPT) were used during the software development process for code generation and debugging assistance. This work is part of RCSB PDB core operations which are jointly funded by National Science Foundation (NSF) (DBI-2321666, PI: S.K. Burley), the US Department of Energy (DE-SC0019749, PI: S.K. Burley), and the National Cancer Institute, the National Institute of Allergy and Infectious Diseases, and the National Institute of General Medical Sciences of the NIH (R01GM157729, PI: S.K. Burley).